\documentclass{pap}

\usepackage{graphicx,isolatin1,psfrag}
\usepackage[hypertex]{hyperref}

\newcommand{\pq}[1]{\left[{#1}\right]}
\newcommand{\p}[1]{\left({#1}\right)}

\newcommand{\D}{\mathrm{d}}
\newcommand{\E}{\mathrm{e}}
\newcommand{\I}{\mathrm{i}}
\newcommand{\FF}{\mathcal{F}}
\newcommand{\PP}{\mathcal{P}}
\newcommand{\DD}{\mathcal{D}}
\newcommand{\LL}{\mathcal{L}}
\newcommand{\HH}{\mathcal{H}}
\newcommand{\NN}{\mathcal{N}}
\newcommand{\kt}{k_\mathrm{B}T}
\newcommand{\average}[1]{\left<{#1}\right>}

\title{Work distribution and path integrals in general mean-field systems}
\shorttitle{Work distribution in general mean-field systems}

\author{A. Imparato\thanks{Associati INFN, Sezione di Napoli.}
\thanks{Corresponding author. Email: \email{imparato@na.infn.it}}
\and L. Peliti $(^{*})$}

\institute{Dipartimento di Scienze Fisiche and Unità INFM,\\
Università ``Federico II'', Complesso Monte S. Angelo, I--80126 Napoli (Italy)}

\begin{document}
\maketitle


\begin{abstract}
We consider a mean-field system described by a general collective
variable $M$, driven out of equilibrium by the manipulation
of a parameter $\mu$. Given a general dynamics compatible with
its equilibrium distribution, we derive the evolution equation
for the joint probability distribution function of $M$ and the
work $W$ done on the system. We solve this equation by path
integrals. We show that the Jarzynski equality
holds identically for these dynamics, both
at the path integral level and for the classical paths which
dominate the expression in the thermodynamic limit. We discuss some
implications of our results.
\end{abstract}

In a class of micromanipulation experiments a system initially
in thermal equilibrium is subject to the external manipulation
of a given parameter $\mu$, and the
work $W$ done on it is evaluated \cite{expts,exp1}. This allows one to gather some
interesting information on its structure, and in particular makes
it possible to evaluate the free energy difference between the
initial and final values of the manipulated parameter, by
means of the Jarzynski equality~\cite{JE}
\begin{equation}
\label{JE:eq}
\average{\E^{-\beta W}}=\E^{-\beta \Delta F}.
\end{equation}
In this expression, $\beta=1/\kt$, where $T$ is the temperature,
the average on the lhs is performed over the realizations of the
process, and $\Delta F$ is the difference between the values
of the equilibrium free energy $F$ for the final and the initial
values of the manipulated parameter $\mu$.

In order to evaluate correctly the lhs of eq.~(\ref{JE:eq}) it
is important to gather information on the probability distribution
function of $W$ for a given protocol $\mu(t)$.
In a previous work~\cite{noi}
we pointed out that a convenient way of achieving this goal is
by considering the joint probability distribution function $\Phi_\sigma(W)$
of $W$ and the microscopic state $\sigma$ of the system.
We considered a comparatively simple system, like a pulled
biomolecule subject to kinetic barriers, and obtained the
corresponding work distributions. It is clear, however, that
one should consider in general more complicated systems, described
by collective coordinates and by more complex stochastic equations
of motions.

In a recent work~\cite{felix} F. Ritort introduced a path integral representation
of the work distribution for a very simple system, namely a collection
of noninteracting spins, and showed how its properties could be derived
by drawing on thermodynamical analogies. Here we wish to generalize both
approaches, by considering a system described by a general collective
coordinate $M$, an equilibrium free energy $\FF_\mu(M)$, depending
on a parameter $\mu$ which is manipulated according to a protocol
$\mu(t)$, $0\le t \le t_\mathrm{f}$,
and a general stochastic evolution described by
a partial differential equation of the form
\begin{equation}
\label{evol:eq}
\frac{\partial P}{\partial t}=\widehat\HH\,P,
\end{equation}
where $P(M,t)$ is the probability distribution function of the
collective variable $M$, and $\widehat\HH$ is a differential
operator, depending on $\mu$, and compatible with an equilibrium
distribution
\begin{equation}
\label{eqdistr:eq}
P_\mathrm{eq}(M;\mu) \propto \E^{-\beta \FF_\mu(M)}.
\end{equation}
We should have, therefore, for a given $\mu$ and any value of $M$,
\begin{equation}
\label{eqmu:eq}
\widehat\HH\,P_\mathrm{eq}(M;\mu)=0.
\end{equation}
During the manipulation of $\mu$, the evolution of the system is
described by a stochastic trajectory $M(t)$, and the work done on the system reads
\begin{equation}
W=\int_0^{t_\mathrm{f}} \D t\,  \dot \mu(t) \, \frac{\partial \FF_\mu(M(t))}{\partial \mu}\, .
\end{equation}
The evolution equation for the joint
probability distribution $\Phi(M,W,t)$ of $M$ and $W$ can be derived by
following the steps of ref.~\cite{noi}:
\begin{equation}
\label{phi:eq}
\frac{\partial \Phi}{\partial t}=\widehat \HH \Phi
-\dot \mu \frac{\partial \FF_\mu}{\partial \mu}\frac{\partial\Phi}{\partial W}.
\end{equation}
A similar equation for the joint probability distribution of the {\it microscopic} state and the work has been derived in ref.
\cite{seif}.
It is actually more convenient to introduce the corresponding
generating function $\Psi(M,\lambda,t)$ for the work distribution:
\begin{equation}
\label{psi:def}
\Psi(M,\lambda,t)=\int \D W\,\E^{-\lambda W}\,\Phi(M,W,t).
\end{equation}
One can readily see that $\Psi$ satisfies the following
evolution equation:
\begin{equation}
\label{psi:eq}
\frac{\partial \Psi}{\partial t}=\widehat \HH \Psi
-\lambda\dot \mu \frac{\partial \FF_\mu}{\partial \mu}\Psi.
\end{equation}
Thus, if $\lambda=\beta$,
\begin{equation}
\label{sol:eq}
\PP(M,t)=\exp\left[-\beta\FF_{\mu(t)}(M)\right]/Z_0,
\end{equation}
(where $Z_0$ is the partition
function of the initial state) solves this equation and
satisfies its initial condition
\begin{equation}
\label{initial:eq}
\Psi(M,\lambda=\beta,0)=\PP(M,0)=\frac{\E^{-\beta \FF_{\mu(0)}}}{Z_0}.
\end{equation}
Thus $\Psi(M,\lambda{=}\beta,t)=\PP(M,t)$,~\cite{jarlez} and
by integrating this function upon $M$
one obtains the Jarzynski equality (\ref{JE:eq}).
We have thus shown that the solution of eq.~(\ref{phi:eq})
satisfies the Jarzynski equality identically.

In order to obtain $\Phi$,
one can attempt to solve directly eq.~(\ref{psi:eq}) by numerical means.
This approach runs into numerical difficulties
when the system size $N$ becomes large, because the
corresponding distributions become narrower and narrower.
Much insight is gained by looking at its path integral
solution in the thermodynamic limit, as pointed out in ref.~\cite{felix}.
By going through the steps explained, e.g., in~\cite{FeynHib},
one can see that, if the differential operator
$\widehat\HH$ is given by
\begin{equation}
\label{diffop:def}
\widehat \HH\cdot{}=\sum_{k=0}^\infty \frac{\partial^k}{\partial M^k}
\left\{g_k(M)\cdot{}\right\},
\end{equation}
one has
\begin{equation}
\label{integral:def}
\Psi(M,\lambda,t) = \int \D M_0
\int_{M(0)=M_0}^{M(t)=M} \DD\gamma\DD M\;
\exp\left\{\int \D t'\,\LL(t')\right\}\;\Psi(M_0,\lambda,0),
\end{equation}
where the Lagrangian $\LL$ is given by
\begin{equation}
\LL(t)=\left.\left(\gamma \dot M+\HH(\gamma,M)
-\lambda \dot \mu \,
\frac{\partial \FF_\mu}{\partial \mu}\right)\right|_{\gamma(t),M(t),\mu(t)}.
\end{equation}
Here $\HH(\gamma,M)$ is given in terms of the differential operator
(\ref{diffop:def}),
\begin{equation}
\label{hamiltonian:eq}
\HH(\gamma,M)=\sum_{k=0}^\infty \gamma^k g_k(M,\mu(t)),
\end{equation}
and the path integral is defined by
\begin{equation}
\int\DD\gamma\DD M =\lim_{{\Delta t\to 0, \NN\to \infty}\atop{ \NN\Delta t=t}}
\prod_{k=1}^\NN\frac{\D\gamma(k\Delta t)\,\D M(k\Delta t)}{2\pi\I}.
\end{equation}
It can be readily seen that this formulation encompasses
that discussed in ref.~\cite{felix}.

We denote by $N$ the size of the system, and set $M=Nm$, $\FF_\mu(M)=Nf_\mu(m)$,
$\HH(\gamma,M)=N H(\gamma,m)$.
As $N\to \infty$, the path integral is dominated by the classical
paths $(\gamma_\mathrm{c}(t),m_\mathrm{c}(t))$, satisfying the equations of motion
\begin{eqnarray}
\label{eqm}
0&=&\dot m+\frac{\partial H}{\partial \gamma};\\
\label{eqgamma}
0&=& -\dot \gamma +\frac{\partial H}{\partial m}
-\lambda \dot \mu \frac{\partial^2 f_\mu}{\partial m\partial \mu}.
\end{eqnarray}
The condition at $t=0$ for these equations of motion
must now be discussed. For all values of $\lambda$, $\Psi(M,\lambda,t{=}0)$
satisfies eq.~(\ref{initial:eq}). Thus the integral over the
variables at $t=0$ takes the form
\begin{equation}
\Psi(M,\lambda,t=0)=\frac{\E^{-\beta \mathcal{F}_{\mu(0)}(M_0)}}{Z_0}=\int \frac{\D \gamma_0\D M'_0}{2\pi\I}\;
\E^{\gamma_0(M_0-M'_0)}\,\frac{\E^{-\beta \mathcal{F}_{\mu(0)}(M'_0)}}{Z_0}.
\end{equation}
Evaluating this integral by the saddle point method one obtains $m'_0=m_0$
and that $\gamma_0$ and $m_0$ satisfy
\begin{equation}
\label{selfcon:eq}
Q\equiv\beta^{-1}\gamma+\frac{\partial f_{\mu}}{\partial m}=0.
\end{equation}
Thus, for any given value of $m_0$ one can evaluate the
corresponding value of $\gamma_0$ by imposing eq.~(\ref{selfcon:eq}) at $t=0$,
and then integrate the equations of motion (\ref{eqm},\ref{eqgamma}) for
$0\le t\le t_\mathrm{f} $. This will yield the values $(m_\mathrm{f},\gamma_\mathrm{f})$
of $m(t_\mathrm{f})$ and $\gamma(t_\mathrm{f})$ respectively.
By evaluating the corresponding action one can thus obtain the
value of $\Psi(N m_\mathrm{f},\lambda,t_\mathrm{f})$.
The generating functional of the work distribution $\Gamma(\lambda,t_\mathrm{f})$ is
given by the integral of $\Psi(N m_\mathrm{f},\lambda,t_\mathrm{f})$ with respect to $m_\mathrm{f}$, see eq. (\ref{psi:def}).
As pointed out in \cite{felix}, this corresponds to evaluating
$\Psi(N m_\mathrm{f},\lambda,t_\mathrm{f})$ for that particular solution of the equations of motion
which satisfies $\gamma_\mathrm{f}=0$. Thus one must solve the
equations of motion (\ref{eqm},\ref{eqgamma}) by imposing the
boundary condition (\ref{selfcon:eq}) at $t=0$ and the
other condition $\gamma(t)=0$ at $t=t_\mathrm{f}$.
This can be done by a shooting method: one solves the equations
as a function of the initial value $m_0$ of $m$ and chooses
that value $m_0^*$ for which $\gamma(t_\mathrm{f})=0$.
(This had not been necessary in ref.~\cite{felix},
since in that case the equation for $\gamma$ did not involve $m$.)
This can
be done more easily if one observes that, for $\lambda=0$, the
solution $\gamma(t)\equiv 0$ is the correct one. One can build on
it to evaluate successively the solution for slowly increasing values of $|\lambda|$.
Alternatively, one may resort to a relaxation method,
which is more convenient for slower manipulation protocols.
Once the generating function $\Gamma(\lambda,t_\mathrm{f})$
of the work distribution is obtained, the
distribution itself can be obtained by a saddle point integration,
which amounts to a Legendre transformation on $\log \Gamma$:
\begin{equation}
P(W,t_\mathrm{f})=\int \D \lambda\,  \E^{\lambda W+ \log \pq{\Gamma (\lambda,t_\mathrm{f})}}= \E^{\lambda^* W+ \log\pq{ \Gamma (\lambda^*,t_\mathrm{f})}}\, ,
\end{equation}
where $\lambda^*$ is implicitly defined by
$\left.\partial_\lambda \log\Gamma\right|_{\lambda^*}=-W$. It is worth noting that we obtain in this way a probability distribution of the work {\it per spin}, $w=W/N$, which has the form $P(w)\propto \exp\pq{N g(w)}$, where $g(w)$ is  a function of $w$ alone.

We now show that the solution obtained in this way satisfies
the Jarzynski equality identically.
We first show that, for $\lambda=\beta$, the solution
of the classical equations of motion (\ref{eqm},\ref{eqgamma})
satisfy an equation analogous to (\ref{selfcon:eq})
at all times. By
multiplying both sides of eq.~(\ref{eqmu:eq}) by
$\E^{-\gamma M}$ and integrating by parts over $M$ one
obtains
\begin{equation}
\label{saddle:eq}
\int \D M\;\HH(\gamma,M)\,\E^{-\beta\FF_\mu(M)-\gamma M}=0,
\end{equation}
where $\HH(\gamma,M)$ is given by (\ref{hamiltonian:eq}).
Evaluating this integral by the saddle point method
in the large $N$ limit, we obtain
\begin{equation}
\label{zeroham:eq}
H(\gamma,m^*)=0,
\end{equation}
if $\gamma$ and $m^*$ are related by
(\ref{selfcon:eq}).
By differentiating eq.~(\ref{zeroham:eq}) with respect to $\gamma$
at fixed $\mu$ we obtain
\begin{equation}
\label{derH:eq}
\frac{\partial H}{\partial\gamma}+\left.\frac{\partial H}{\partial m}\right|_{m^*}
\left.\frac{\partial m^*}{\partial \gamma}\right)_\mu=0.
\end{equation}
Let us also take the derivative of eq.~(\ref{selfcon:eq}) with
respect to $\gamma$ at fixed $\mu$, we obtain
\begin{equation}
\label{dergamma:eq}
-\beta^{-1}=\frac{\partial^2 f_\mu}{\partial m^2}
\left.\frac{\partial m^*}{\partial \gamma}\right)_\mu.
\end{equation}
By multiplying both sides of eq.~(\ref{derH:eq}) by  $\partial^2 f_\mu/\partial m^2$ and substituting eq.~(\ref{dergamma:eq}), we obtain the following relation
\begin{equation}
\frac{\partial^2 f_\mu}{\partial m^2}\frac{\partial H}{\partial \gamma}-\beta^{-1}\frac{\partial H}{\partial m}=0,
\label{eqhf}
\end{equation}
which holds when $\gamma$ and $m$ are related by eq.~(\ref{selfcon:eq}).
We can now evaluate the time derivative of the lhs of
 eq.~(\ref{selfcon:eq}), when  $\gamma$ and $m$ satisfy eqs.~(\ref{eqm},\ref{eqgamma}). We have
\begin{equation}
\dot Q = -\beta^{-1} \dot \gamma -\frac{\partial^2 f_\mu}{\partial m^2} \dot m -\frac{\partial^2 f_\mu}{\partial m\partial \mu}\dot \mu
= -\beta^{-1}\p{\frac{\partial H}{\partial m}-\beta \frac{\partial^2 f_\mu}{\partial m\partial \mu}\dot \mu }+\frac{\partial^2 f_\mu}{\partial m^2} \frac{\partial H}{\partial \gamma}- \frac{\partial^2 f_\mu}{\partial m\partial \mu}\dot \mu\, .
\end{equation}
The second and the last term cancel out. Substituting eq.~(\ref{eqhf}), we see that also the first and the third therm cancel out. Thus if $\gamma$ and $m$ satisfy eqs.~(\ref{eqm},\ref{eqgamma}) at all times, and satisfy eq. (\ref{selfcon:eq}) at a given time, they satisfy this last equation at any time.

Thus, for $\lambda=\beta$, the Lagrangian, evaluated along the classical
path, is given by
\begin{equation}
\LL_\mathrm{c}=N\left[\gamma\dot m-\beta\dot \mu \frac{\partial f_{\mu}}{\partial \mu}\right]
=N\left[-\beta \frac{\partial f_\mu}{\partial m}\dot m-\beta \frac{\partial f_\mu}{\partial \mu}\dot\mu\right]
=-\beta N \frac{\D f_\mu}{\D t},
\end{equation}
where we have exploited eq.~(\ref{selfcon:eq}). Substituting
this expression in eq.~(\ref{integral:def}) one recovers eq.~(\ref{sol:eq}) and
the Jarzynski equality.

\begin{figure}
\psfrag{w}{$w$}
\psfrag{p(w)}{$P(w),\Omega(w)$}
 \onefigure[width=8cm]{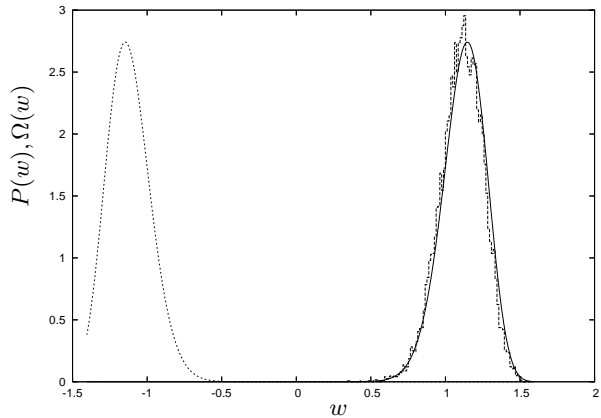}%
 \caption{Results for the system described by
the differential operator (\ref{FP}) with equilibrium free energy
(\ref{freeen:eq}), manipulated according to the
protocol (\ref{manh:eq}). Continuous
line: probability density $P(w)$ of the work ``per spin'' $w=W/N$, with N=50.
Dashed line: histogram of the work obtained by 10000 simulations
of the process via the Heun algorithm~\cite{Greiner}. Dotted line:
$\Omega(w)=P(w)\exp(-\beta N w)$, whose
integral verifies the Jarzynski equality. One can see that
there are no simulation points in the range in which $\Omega(w)$ is essentially different from zero.
Thus, in this case, a quantitative verification of the Jarzynski equality
would be practically impossible.
\label{results:fig}}
 \end{figure}

In the following we consider a simple mean-field Ising-like
system, with free energy
\begin{equation}
\label{freeen:eq}
\FF(M)=-\frac{J}{2 N}M^2-hM-TS(M),
\end{equation}
where
\begin{equation}
S(M)=Nk_\mathrm{B}\left[\ln 2-\frac{1}{2}\ln(1-m^2)-
\frac{m}{2}\ln\left(\frac{1+m}{1-m}\right)\right].
\end{equation}
The evolution of the system is governed by a Fokker-Planck
equation with differential operator
\begin{equation}
\label{FP}
\widehat \HH \cdot {}=\frac{\partial}{\partial M}\left[
\omega_0\left(\frac{\partial\FF }{\partial M}\right)\cdot{}
+\beta^{-1}\omega_0\frac{\partial}{\partial M}\cdot{}\right].
\end{equation}
We show in fig.~\ref{results:fig} the results of our calculation
for this system, with $\beta=1$, $J=0.5$, $\omega_0 \equiv 1$, $N=50$,
and where the external field $h$ is manipulated according to the
protocol
\begin{equation}
\label{manh:eq}
h(t)=h_0+(h_1-h_0)\frac{t}{t_\mathrm{f}}; \qquad 0<t<t_\mathrm{f}.
\end{equation}
We have set $h_0=-h_1=-1$, $t_\mathrm{f}=2$.
We compare the predictions with the results
of the simulation of the Langevin equation underlying
eq.~(\ref{FP}), integrated according
to the Heun algorithm~\cite{Greiner}.
We see that in this
case the agreement between the simulation results and the
distribution obtained by our analysis is quite good: nevertheless,
one can also conclude that even for such a small system
the possibility of using the results of the nonequilibrium
manipulation protocol to obtain information on the equilibrium
free energy is rather remote. Indeed, we have plotted
in fig.\ref{results:fig} the quantity $\Omega(w)=P(w)\E^{-\beta N w}$,
whose integral should give $\E^{-\beta \Delta F}$ (which is
equal to 1 in our case). We can see that not a single point
among the simulated 10000 falls in the  range in which $\Omega(w)$
is essentially different from 0.

\begin{figure}
\psfrag{w}{$w$}
\psfrag{p(w)}{$P(w)$}
\onefigure[width=8cm]{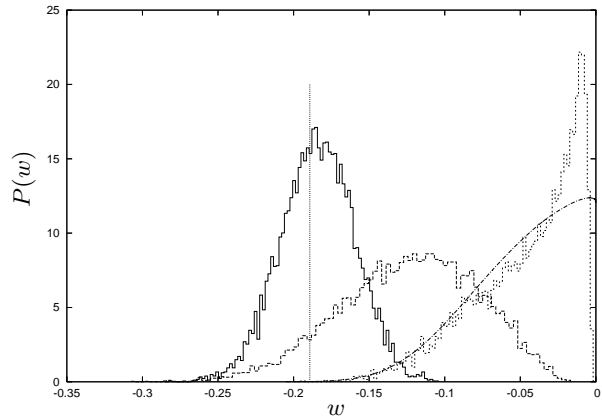}%
\caption{%
Results for the system described by
the differential operator (\ref{FP}) with equilibrium free energy
(\ref{freeen:eq}), manipulated according to the
protocol (\ref{manj:eq}).
Histograms over 10000 simulations for each value of $N$.
Continuous line: $N=1$. Dashed line: $N=10$. Dotted line $N=100$.
Dash-dotted line: Calculated $P(w)$ for $N=100$.
The calculated $P(w)$ (not shown) for smaller values of $N$
have the same general behavior as for $N=100$, but
are wider.
Vertical line: Thermodynamic value of the work
$w_\mathrm{rev}=\Delta F/N$.
\label{jj:fig}}
\end{figure}

When $h=0$ and $\beta J>1$, our system undergoes a symmetry-breaking
phase transition. Now assume that $h=0$, $\beta J(0)<1$,
$\beta J(t_\mathrm{f})>1$. Since $\gamma(t_\mathrm{f})=0$, the identically
vanishing solution $(\gamma_\mathrm{c}\equiv 0,m_\mathrm{c}\equiv 0)$
satisfies the equations of motion (\ref{eqm},\ref{eqgamma}) and
the initial condition. This would lead to $W\equiv 0$ in contradiction
with the Jarzynski equality. However, this solution does not
correspond to the extremum of the exponent in eq.~(\ref{saddle:eq}),
which is given by the symmetry-breaking solution. It is therefore
necessary to perform a quasi-average, by introducing a small
symmetry-breaking field $h$, evaluating the classical action,
and going to the limit $h\to 0$ at the end of the calculation.
We show in fig.~\ref{jj:fig} the results of this
calculation. We have set $h_0=h_1=0.01$, and manipulated $J$
according to the protocol
\begin{equation}
\label{manj:eq}
J(t)=J_0+(J_1-J_0)\frac{t}{t_\mathrm{f}},\qquad 0<t<t_\mathrm{f},
\end{equation}
with $J_0=0.5$, $J_1=1.5$, $t_\mathrm{f}=10$, $\beta=1$.
Notice that for such a protocol the work done on the system is non-positive, since $w=-\int \D t \dot J(t) m^2(t)$.
The figure contains the results of 10000 simulations each
with $N=1$ (continuous line), $N=10$ (dashed line), $N=100$
(dotted line), and the expectations of the theory for
$N=100$ (dash-dotted line). The agreement between theory and simulations
is poor for the smaller values of $N$, reasonable for $N=100$,
and improves as $N$ grows (other data not shown). However, notice
that for small values of $N$ the simulated values of $w$
are consistently smaller than those obtained asymptotically for large $N$,
and lie
closer to the thermodynamical value $w_\mathrm{rev}$
denoted by a vertical line. Indeed, for small values of $h$ and
large values of $N$ the mean-field system has difficulties
in keeping close to the true thermodynamic minimum, whereas a smaller
system can reach it by activated processes.

We have derived the differential equation (\ref{phi:eq}) for the
joint probability distribution of the collective coordinate
$M$ and the work $W$ for a system evolving according to a general
dynamics. We have shown that
the solution of this equation satisfies the
Jarzynski equality identically.
We find that this equation can be solved by the
path integral (\ref{integral:def}).
This solution
can be evaluated, asymptotically for large system
sizes $N$, by a saddle-point integration, leading
to a two-point boundary value problem
for the solution of the classical equations of motions,
thus generalizing the approach of ref.~\cite{felix}
to interacting systems.
We have seen that the Jarzynski equality can be
effectively exploited to obtain the free energy
difference only for very small systems.
We have pointed out that
the method has to be handled carefully if the manipulation
protocol leads in the vicinity of a phase transition.
(The situation is even more delicate if the protocol
crosses a discontinuous transition line.)
It is possible to extend the approach to disordered
mean-field systems like the Hopfield model.
By allowing
one to explore the work probability distribution function,
the formalism can be of use in identifying the
manipulation protocol most likely to yield the desired results.

\begin{acknowledgments}
We are grateful to Felix Ritort for a critical reading of the manuscript.
LP is grateful to Silvio Franz for thought-provoking remarks.
\end{acknowledgments}

\end{document}